\documentstyle[hip-artc]{article}  
\volnumber{10}  \edyear{1999}  \frompage{345} \topage{349}                %| 
\recrevdate{1 January 1999}                                              %| 
\def\rightmark{The strong coupling constant} 
\def\leftmark{E. Recami et al.}
\title{The strong coupling constant} 
\authors{
{\twerm Erasmo Recami $^1$, Tonin-Zanchin$^{2}$, Antonino Del Popolo$^{3}$ and Mario Gambera$^{3}$ %
}\\[2.812mm]
{\normalsize
\hspace*{-8pt}$^1$ Facolt\`{a} di Ingegneria, Universit\`{a}
Statale di Bergamo, \\ 
24044--Dalmine (BG), Italy\\[0.2ex] 
\hspace*{-8pt}$^2$ Dept. of Applied Mathematics, State University at
Campinas,  \\  
S.P., Brasil\\[0.2ex]
\hspace*{-8pt}$^3$ Dipartimento di Matematica, Universit\'a Statale di Bergamo,  \\  
24129--Bergamo, Italy\\[0.2ex]
}}
\abstract{In this paper we fix our attention, on hadron structure, and show
that also the strong interaction strength
$\alpha_{\rm S}$, ordinarily called the ``{\it (perturbative)
coupling--constant square}", can be evaluated within our theory, and found to decrease (increase)
as the ``distance" $r$ decreases (increases).  This yields both the 
confinement of the hadron constituents, and their
asymptotic freedom:  in 
qualitative agreement with the experimental evidence.  In other terms, our
approach leads us, on a purely theoretical ground, to a dependence of
$\alpha_{\rm S}$ on $r$ which had been previously found only on 
phenomenological and heuristic grounds. \  We expect the above agreement 
to be also quantitative, on the basis of a
few checks performed in this work.}
\PACS{$^b$}
 
\begin{document}
 
\maketitle

\section{Introduction}
In the past years, a unified approach to strong and
gravitational interactions was proposed (\cite{bib1,bib2,bib3,bib4}), which 
used the geometrical
methods of general relativity; and assumed covariance of physical laws
under global {\it discrete} dilations.  It yielded results similar to those 
given by  the
``strong gravity" theory (\cite{bib5,bib6}).\\
Within such an approach, and in connection with hadron structure, we 
came in particular to associate hadron constituents with
suitable stationary, axisymmetric solutions of certain new
Einstein--type equations, supposed to describe the strong field
inside hadrons. \  Those Einstein--type equations are nothing but the
ordinary Einstein equations (with cosmological term) suitably
scaled down (\cite{bib3,bib4}). As a consequence, the cosmological constant
$\Lambda$ and the gravitation universal constant $G$ (or the masses $M$) 
result, in our theory, to be scaled up
and transformed into a ``hadronic constant" $\lambda$ and into a ``strong
universal constant" $N$ (or into ``strong masses" $g$),
respectively (\cite{bib1,bib2,bib3,bib4,bib7,bib8}). Our field equations, to be valid inside a hadron,
are therefore:
 \begin{equation}
R_{\mu \nu} - \frac{1}{2} g_{\mu \nu} R^{\rho}_{\rho} + \lambda
g_{\mu \nu} = - KNT_{\mu \nu} \; ; \;\;\; [K \equiv \frac{8 \pi}{c^{4}}] ,
\end{equation}
where, because of simple dimensional considerations$^{[2]}$, 
$$\lambda \simeq {\rho_{1}}^{2}\Lambda \;;\;\;\; N \equiv \rho G \;;\;\;\;
\rho \simeq \rho_{1} \simeq 10^{41} \; .$$
If we adopt  $\Lambda 
\simeq 10^{-52} {\rm m}^{-2}$, then we get for the ``strong cosmological 
constant" (hadronic constant) the value $\lambda \simeq 10^{30} {\rm m}^{-2} = 
(1  \; \; {\rm fm})^{-2}$. \\
Throughout this paper, we shall choose 
the signature  $-2$. When convenient, we shall use units such that it be 
also $c = 1$. The {\em simplest} solution of Eqs.(1) is the Schwarzschild--de Sitter's,
corresponding to the metric generated by a central, static, 
spherically--symmetric distribution of strong charge; {\em i.e.}, to the
metric generated by a hadron constituent (say, a quark) when neglecting 
its electric charge and intrinsic angular momentum:
\begin{eqnarray}
ds^{2} & \equiv & g_{\mu \nu}{\rm d}x^{\mu}{\rm d}x^{\nu}  = \nonumber\\
& = & (1-\frac{2Ng_{\rm o}}{r}-\frac{\lambda r^{2}}{3}){\rm d}t^{2} -
(1-\frac{2Ng_{\rm o}}{r}-\frac{\lambda r^{2}}{3})^{-1}{\rm d}r^{2} -
r^{2}({\rm sin}^{2}\theta {\rm d}\varphi^{2}+{\rm d}\theta^{2})
\end{eqnarray}
where $g_{\rm o}$ is the strong charge of the considered constituent, and 
($t,r,\theta,\varphi$) are spherical (Schwarzschild--type) coordinates. Let
us stress once more that, in the present units, $g_{\rm o}$ is equal to the
rest--mass $M_{\rm o}$ of the hadron constituent.
\section{Strong--charge and its dependence on $r$}
Now, we consider
the geodesic motion of a
test--particle in the metric (2).
 Our test particle, when 
free--falling, will be endowed (\cite{bib1}) with a constant total--energy $E_{\rm o} 
\equiv g'_{\rm o} c^{2}$, which in the previous coordinates can be written
\begin{equation}
E_{\rm o} \equiv g'_{\rm o} c^{2} = g_{tt} p^{t}\; ; \;\;\;\;\;\;
\end{equation}
where $ p^{t} \equiv p^{\rm o} \equiv g'_{\rm o} \;
{\frac{{\rm d}t}{{\rm d}s}} $ and $g_{tt} \equiv g_{\rm oo}$, and $g'_{\rm o}$ is the ({\em rest}) 
strong--mass of the test particle.\\
Since the Schwarzschild--type coordinates do not correspond to any
physical observer, let us pass ---however--- to the {\em local coordinates}
($T,R, \theta , \varphi$), associated with observers {\em at rest} w.r.t. [with
respect to] the metric at each point ($r, \theta , \varphi$) of the 
considered space:
\
$${\rm d}T \equiv \sqrt{g_{tt}} {\rm d}t \; ; \;\;\;\;\;\; {\rm d}R \equiv
\sqrt{-g_{rr}} {\rm d}r \; ,$$
\
where $g_{rr} \equiv g_{11}$.  \  The local observers measure a new 
total--energy $E_{\rm \ell}$ for the considered test particle, where
$E_{\rm \ell}$ is no longer a constant of the motion and is related to
$E_{\rm o}$ by
%\vspace*{0.5 cm}
\begin{equation}
%\hfill{$
E_{\rm \ell} \equiv g'_{\rm o} \frac{{\rm d}T}{{\rm d}s}
\equiv p^{T} = \sqrt{g_{tt}} p^{t}   ,
%\hfill} 
\end{equation}
%(4a)
%\vspace*{0.5 cm}
that is 
%\vspace*{0.5 cm}
%\hfill{$
\begin{equation}
E_{\rm \ell} \equiv g' c^{2} = \frac{g'_{\rm o} 
c^{2}}{\sqrt{g_{tt}}} \; 
%.$
\end{equation}
%\hfill} (4b)
%\vspace*{0.5 cm}
In the static case \  $\sqrt{g_{tt}} = 
\sqrt{1-V^{2}}$, \  provided that $V$ is measured by the local observers. 
\
\
The physical meaning of Eqs.(4)-(5) is more evident if, instead of
setting $M \equiv g$ and $N = \rho G$, we put $N = G = 1$ so that
(in such new units) for the strong charge it holds
%\vspace*{0.5 cm}
%\hfill{$
\begin{equation}
g = \sqrt{\rho}M \; ; \;\;\; g_{\rm o} = \sqrt{\rho}M_{\rm o} 
%.$
\end{equation}
%\hfill} (5)
%\vspace*{0.5 cm}
Here, let us remind that,
for $M \simeq m_{\pi}$, one gets $g =$ Planck--mass; \ that is, 
the strength of the interaction between two (strongly interacting)
quarks is equal to the strength of the interaction between two
(gravitationally interacting) particles endowed with the Planck mass.   
\
In whatever units, Eqs.(4)-(5) tell us that 
the strong charge $g'$ of the test--particle does change with
its speed $V$, w.r.t. the local observers, as follows:
%\vspace*{0.5 cm}
%\hfill{$
\begin{equation}
g' = \frac{g'_{\rm o}}{\sqrt{1-V^{2}}} \; ;
\end{equation}
%$
%\hfill} (6a)
%\vspace*{0.5 cm}
where $V \equiv {{\rm d}R}/{{\rm d}T}$  \  (and $g'$) are measured 
in the local reference--frames: actually, it is in these 
frames that they have a {\em direct} physical meaning.$^{[7]}$ \  In the 
case of generic motion, we are left, of course, with the relation
%\vspace*{0.5 cm}
%\hfill{$
\begin{equation}
g' = \frac{g'_{\rm o}}{\sqrt{g_{tt}}} \; 
%.$
\end{equation}
%\hfill} (6b)
%\vspace*{0.5 cm}
that is  the strong charge (or strong mass) of a particle does
depend, inside a hadron, on the particle speed exactly as the ordinary 
gravitational mass does in our space-time.\\
Notice that Eqs. (7)-(8) allow us to express the value of the strong charge $g'$ 
as a function, {\em e.g.}, of its radial coordinate $r$ relative 
to the source--quark.  Namely, in the case of Eq. (7) one has
$$V^{2} = 2Ng_{\rm o}/r + \lambda r^{2}/3 \; ,$$
and therefore from Eq. (7) one gets: 
\begin{equation}
g' = \frac{g'_{\rm o}}{\sqrt{1-2Ng_{\rm o}/r-\lambda
r^{2}/3}} \; 
\end{equation}
\section{The strong coupling constant}
Similarly with the electromagnetic case, in which $\alpha_{\rm E} = (ke^{2})/(\hbar c)$, the
strong interaction strength is defined (\cite{bib1,bib2,bib3,bib4,bib7,bib8}) as
\
$$\alpha_{\rm S} \equiv \frac{{Ng'}^{2}}{\hbar c}$$
\
which is also a pure number and ---passing to the field-theoretical 
language--- corresponds to the dimensionless square of the vertex
coupling--constant.  Let us recall that $g'$ is measured
by the local observer.   From Eqs.(9) we obtain 
\begin{equation}
\alpha_{\rm S} = (N/\hbar c){g'_{\rm o}}^{2}(1-2Ng_{\rm o}/r-
\lambda r^{2}/3)^{-1}, 
\end{equation}  
where $g_{\rm o}$, $g'_{\rm o}$ are the {\it
rest} strong mass of the 
source--quark and the test--constituent, respectively.  In the case when also 
$g'$ is a quark, we have:
\begin{equation}
\alpha_{\rm S} \simeq \frac{N}{\hbar c} \; \; \frac{{g_{\rm o}}^{2}}
{1-2Ng_{\rm o}/c^{2} r - \lambda r^{2}/3} \; 
\end{equation}
\
Therefore, the strong interaction strength $\alpha_{\rm S}$, which in
elementary particle physics is ordinarily called the ``(perturbative)
coupling--constant square", is predicted by our approach to decrease 
(increase) as the ``distance" $r$ decreases (increases).  This yields
both the {\em confinement} of the constituents (for values of $r
\sim 1$ fm), and their so--called {\em asymptotic freedom}: in
qualitative agreement with the experimental evidence.  In other words,
our approach leads us ---on purely theoretical grounds--- to a dependence
of $\alpha_{\rm S}$ on $r$ which was previously found, within the 
perturbative QCD$^{[8]}$, only on phenomenological and heuristic
grounds.\\
When performing explicit calculations to evaluate $\rho$, at the beginning we
tacitly compared (\cite{bib1}) the gravitational interaction strength 
$G{m_{\rm o}}^{2}$/$\hbar c$
with the value $N{g_{\rm o}}^{2}/\hbar c \simeq 14$ corresponding to the 
$pp\pi$
coupling constant square. However, the gravitational interaction strength
should be compared with the analogous strength for the interaction between two
small {\em components} of the corresponding ("reference") hadron, or rather
of a constituent quark of its. Such a strength is unknown. We know, however,
the quark--quark--gluon coupling constant square (\cite{bib10}) for the simplest
hadrons: $N{g_{\rm o}}^{2}/\hbar c \simeq 0.2$, then the best 
value of $\rho$ that we can work out, for calculations inside such hadrons, is 
$\rho \simeq 10^{38} \div 10^{39}$. \\
We, now, explicit the dependence of $g'$ on the radial
coordinate $r$, by expressing $V \equiv {\rm d}R$/${\rm d}T$ as a function
of $r$ starting directly from the geodesic equation, since in our metric
the geodesic motion is always a motion in a plane, we fix $\theta = \pi$/$2$
and then from the geodesic equation one gets
$$({\rm d}r/{\rm d}s)^{2} = 1/H^{2} - (1-2Ng_{\rm o}/r-\lambda r^{2}/3)
(1+a^{2}/r^{2}) \; ,$$
where $1/H$ and $a$ are rest-energy and angular momentum, respectively, for
unit rest-mass. The last equation yields
\begin{equation}
V^{2} \equiv (\frac{{\rm d}R}{{\rm d}T})^{2} = 1 - H^{2}(1-\frac{2Ng_{\rm o}}
{r}-\frac{\lambda r^{2}}{3}) \; .
\end{equation}
We observe that, for $\lambda > 0$, the minimum of $V^{2}$ is got for $r = 
(3Ng_{\rm o}/\lambda)^{1/3}$.\\
Now, we consider Eq.(12) in connection with Eqs.(7)-(8)-(9), and let us here
emphasize that  for
$\rho_{1} = 10^{41}$; $\rho = 10^{38}$, and $g_{\rm o} = 
m_{\rm p}$/$3 \simeq 313$ MeV/$c^{2}$]  the minimum of $g'$, namely  \
$g' \simeq 1.2 {g'}_{\rm o}$, \  
is obtained at $r \simeq 0.6 \;\; {\rm fm}$.\\
\vspace*{0.6 cm}
\section{Conclusion}
We have seen that the strong interaction strength,
$\alpha_{\rm S}$, ordinarily called the ``(perturbative) coupling--constant 
square", can be evaluated within our theory, and found to decrease 
(increase) as the ``distance" $r$ decreases (increases).  This yielded
the confinement of the constituents (for large values of $r$),
as well as their asymptotic freedom
(for small values of $r$ inside the hadron): in
qualitative agreement with the experimental evidence.  In other words
our approach led us, on a purely theoretical 
ground, to a dependence
of $\alpha_{\rm S}$ on $r$ which had previously been found  
only on phenomenological and heuristic
grounds.\\
~\\

\section*{Acknowledgement(s)}
Useful discussions are acknowledged with
V. de Sabbata, R.H.A.
Farias, R. Garattini, E. Giannetto, A. Insolia, L. Lo Monaco,
G.D. Maccarrone,
E. Majorana jr., R.L. Monaco, E.C. de Oliveira, N. Paver, M. Pav\v{s}i\v{c},
F. Raciti, G. Salesi,
S. Sambataro, P. Saurgnani, C. Sivaram, G. Tagliaferri,
M.T. de Vasconselos, J. Vaz
and particularly with  P. Ammiraju, P. Bandyopadhyay,
L.A. Brasca--Annes, A. Bugini, A. Italiano, L. Mandelli, A. van der Merwe,
W.A. Rodrigues Jr., J.A. Roversi, P. Srivastava.
Very special thanks go to Yuval Ne'eman.

\end{document}